\documentclass[sn-mathphys-num,referee]{sn-jnl}
\usepackage{graphicx}
\usepackage{amsmath}

\begin{document}

\title{Efficient stochastic simulation of piecewise-deterministic Markov processes and its application to the Morris-Lecar model of neural dynamics
}

\author{\fnm{Arkady} \sur{Pikovsky}}\email{arkady.pikovsky@gmail.com}
\affil{\orgdiv{Institute for Physics and Astronomy}, \orgname{University of Potsdam}, \orgaddress{\street{Karl-Liebknecht-Str. 24-25}, \city{Potsdam}, \postcode{14476}, \country{Germany}}}

\abstract{
Piecewise-deterministic Markov processes combine continuous in time dynamics with jump events, the rates of which generally depend on the continuous variables and thus are not constants. This leads to a problem in a 
Monte-Carlo simulation of such a system, where, at each step,  one must find the time instant of the next event. The latter is determined by an integral equation and usually is rather slow in numerical implementation. We suggest a reformulation of the next event problem as an ordinary differential equation where the independent variable is not the time but the cumulative rate. This reformulation is similar to the H\'enon approach to efficiently constructing the Poincar\'e map in deterministic dynamics. The problem is then reduced to a standard numerical task of solving a system of ordinary differential equations with given initial conditions on a prescribed interval. We illustrate the method with a stochastic Morris-Lecar model of neuron spiking with stochasticity in the opening and closing of voltage-gated ion channels.}  

\maketitle

\section{Introduction}
\label{intro}

Piecewise-deterministic Markov processes (PDMPs) are a broad class of stochastic processes with many applications.
Mathematical foundations and properties can be found in books \cite{davis2018markov,jacobsen2006point}.
Sometimes, PDMPs are called hybrid stochastic systems~\cite{li2017review,bressloff2018stochastic,hespanha2018stochastic}. Roughly speaking, PDMP is a generalization of a standard Markov process, which consists of jumps at random instants of time, to a situation where there is also deterministic evolution of some variables between the jumps. The classic example of a PMDP is stochastic neuron dynamics~\cite{hille2001}. Here, the membrane voltage is a continuous variable that varies deterministically according to the capacitance discharge equation. The conductances of the ion channels are random because these channels can spontaneously open and close, and this is modeled with Markov processes. The random and the deterministic dynamics depend mutually on each other: the voltage discharge depends on random conductances, and the rates according to which the channels open and close depend on the voltage.
Another example is a gene regulatory network~\cite{kurasov2018stochastic}. Here, concentrations of the proteins are continuously varying variables, while the activation state of the genes follows the Markovian jump process. 
PDMPs also appear in the description of intracellular transport, where motor cargos can randomly switch between motility states~\cite{bressloff2018stochastic}. We also mention applications in insurance risk modeling~\cite{Embrechts_Schmidli_1994}, communication networks~\cite{hespanha2005model}, DNA replication~\cite{lygeros2008stochastic}, and the individual human behavior models~\cite{Hawker-Siekmann-24}.

The essential property of PDMPs is that the deterministic and stochastic parts are \textit{interdependent}. The states that vary at the Markov discrete jumps influence the deterministic dynamics, and the latter influences the Markov jumps via the variation of the jump rates. Of course, there are simpler situations where the influence is only in one direction. For example, if the rates do not depend on the continuously varying states, one has a model of Markovian noise driving; in the simplest case, it reduces to a dichotomic noise~\cite{hl1984}. Close to the PDMPs are 
Markov processes with prescribed time-dependent rates. This case will also be included below.

As it is already apparent from the description above, the main challenge in the numerical simulation of a PDMP is finding the jump event times; all other ingredients are straightforward. Thanks to D. T. Gillespie~\cite{gillespie1977exact}, there is an exact algorithm for simulation of the Markov process trajectories; see also recent advances in~\cite{wilkinson2018stochastic,masuda2022gillespie}. However, a numerical difficulty appears if the rates are not constant. While for continuous rates, one defines the interval between jumps from an easily generated random number with exponential distribution, in the case of non-stationary rates, one must solve an integral equation--precisely this difficulty we address here below. We will demonstrate how to reformulate the problem so that finding the inter-jump interval reduces to a standard initial-value problem for a system of ordinary differential equations (ODEs).

The paper is organized as follows. In Section~\ref{sec:fm}, we formulate the basic model. Section~\ref{sec:ssm} describes the efficient approach for stochastic simulations. In Section~\ref{sec:ml}, we demonstrate it for the Morris-Lecar model of neuron dynamics. For this example, we elaborate on the accuracy and the performance of the method compared to the other approaches in Section~\ref{sec:ap}. We conclude  in Section~\ref{sec:concl}.

\section{Formulation of the model and its simulation with the Gillespie direct method}
\label{sec:fm}

Here, we formulate a rather generic PDMP. The dynamics consists of purely deterministic evolution epochs interrupted by discrete jump events. There is a set of  variables $\vec{X}(t)$ that evolve during deterministic epochs according to ODEs
\begin{equation}
\frac{d \vec{X}}{dt}=\vec{F}(\vec{X},\vec{Y},t)\;.
\label{eq:bode}
\end{equation}
There may exist another set of variables $\vec{Y}$, which vary only at jump events and
remain constant during deterministic evolution (\ref{eq:bode}). Variables $\vec{X}$ can also vary at jump events. The variables $\vec{X}$ are continuous, while the variables $\vec{Y}$ can be continuous or discrete.
For simplicity, we call below the variables $\vec{X}$ ``continuous'', and the variables $\vec{Y}$ ``discrete''. 

There are generally $M$ different types of discrete events, which are assumed all to be independent Markovian processes with the rates
\begin{equation}
\lambda_i(\vec{X},\vec{Y},t),\quad i=1,\ldots, M\;.
\label{eq:brates}
\end{equation}
Namely, an event $i$ occurs within a small time interval $(t,t+dt)$ with probability $\lambda_i(\vec{X}(t),\vec{Y}(t),t)\,dt$.
If an event happens, generally, all dynamical variables $\vec{X},\vec{Y}$ are transformed according to deterministic or probabilistic rules. However, in some applications, only discrete variables vary at the jumps. 
We will assume that these transformations can be easily implemented in numerical simulations.
Also, the evolution problem (\ref{eq:bode}) is a standard numerical task of solving a system of ODEs, provided the r.h.s. is smooth enough. Usually, it is accomplished with a variant of the Runge-Kutta method.

The main challenge in the numerical simulations is modeling the discrete jump times. Among different methods,
the Gillespie Direct Method (GDM) is one of the most popular (also, the first process method and the next process method are discussed in the literature; a generalization of our approach to these variants is straightforward). 
We present it for the problem formulated above, following Refs.~\cite{alfonsi2005adaptive,anderson2015stochastic},
see also~\cite{anderson2007modified,riedler2013almost}.

In GDM, one uses the independence of different event types and defines the total rate as
\begin{equation}
\Lambda(\vec{X}(t),\vec{Y}(t),t)=\sum_{i=1}^M \lambda_i(\vec{X}(t),\vec{Y}(t),t)\;.
\label{eq:trate}
\end{equation} 
According to this rate, if the last event was at time instant $t^{last}$, then 
the probability of having no event in the time interval $t^{last} <t<T$ is 
\begin{equation}
1-\exp\left(-\int_{t^{last}}^T \Lambda(\vec{X}(s),\vec{Y}(s),s)\;ds\right)\;.
\label{eq:cumprob}
\end{equation}
Note that here the discrete variables $\vec{Y}$ are constants, so that $\vec{Y}(s)=\vec{Y}(t^{last})$.
The way to sample the time instant $t^{next}$ of the next event is first to generate  $r_1$ 
as a random number distributed on the unit interval $0<r_1\leq 1$, and then to calculate $\Delta=-\ln(r_1)$. 
Then, the sampled time of the next event is found from the condition
\begin{equation}
\int_{t^{last}}^{t^{next}} \Lambda(\vec{X}(s),\vec{Y}(s),s)\;ds=\Delta\;.
\label{eq:intcond}
\end{equation}
Because $\vec{X}(t)$ is a solution of the ODE (\ref{eq:bode}), 
it is convenient to reformulate (\ref{eq:intcond}) as an ODE
\begin{equation}
\frac{d\Phi}{dt}=\Lambda(\vec{X}(t),\vec{Y}(t),t)
\label{eq:lambdaode}
\end{equation}
with initial condition $\Phi(t^{last})=0$. The time of the next event is found from the condition
$\Phi(t^{next})=\Delta$. Finding $t^{next}$ is most challenging. As the author of \cite{anderson2007modified} formulated,   ``solving equation (\ref{eq:intcond})
either analytically or numerically will be extremely difficult and time-consuming in all, but
the simplest of cases''. 

After the time instant $t^{next}$ is found, the next standard step in the GDM is performing a jump. Which of possible $M$ jumps is performed requires a further random choice, now from a discrete distribution with probabilities
\begin{equation}
p_i=\frac{\lambda_i(\vec{X}(t^{next}),\vec{Y}(t^{last}),t^{next})}{\Lambda(\vec{X}(t^{next}),\vec{Y}(t^{last}),t^{next})}\;.
\label{eq:dprob}
\end{equation}
This choice is accomplished by generating another random number $r_2$, uniformly distributed on the unit interval 
$0\leq r_2\leq 1$, and finding such integer $j$ that 
\begin{equation}
\sum_{i=1}^{j-1} p_i < r_2\leq \sum_{i=1}^j p_i\;.
\label{jsel}
\end{equation}
Then, the jump of type $j$ is performed, during which variables $\vec{X},\vec{Y}$ are transformed according to the jump rules. Then, the simulation step is accomplished, and one proceeds to the next step. 

\section{Stochastic simulation method}
\label{sec:ssm}

Here, we present an efficient solution to the task of finding the next event from the system (\ref{eq:bode},\ref{eq:lambdaode}). This method is analogous to the M. H\'enon technique for numerical computation of Poincar\'e map in deterministic dynamics \cite{Henon-82}.
In the context of hybrid non-stochastic systems, the H\'enon technique was discussed in \cite{korner2015mathematical}.

Let us first consider the case where the total instantaneous rate $\Lambda$ is bounded from below with a positive constant. This means that, according to (\ref{eq:lambdaode}), the variable $\Phi$ is a strictly monotonically growing function of time. This allows for replacing in the system of ODEs (\ref{eq:bode},\ref{eq:lambdaode}) the independent variable $t$ by $\Phi$. The resulting equations read
\begin{align}
\frac{d \vec{X}}{d\Phi}&=\frac{\vec{F}(\vec{X}(t),\vec{Y}(t),t)}{\Lambda(\vec{X}(t),\vec{Y}(t),t)} \;,\label{eq:newx}\\
\frac{dt}{d\Phi}&=\frac{1}{\Lambda(\vec{X}(t),\vec{Y}(t),t)}\;. \label{eq:newt}
\end{align} 
For the system (\ref{eq:newx},\ref{eq:newt}) the initial condition at $\Phi=0$ is $\vec{X}=\vec{X}(t^{last})$,
$t=t^{last}$. The integration of (\ref{eq:newx},\ref{eq:newt}) should be performed on the \textit{prescribed interval of the independent variable} $0\leq\Phi\leq \Delta$. This is a standard task for ODEs and is usually accomplished with a Runge-Kutta method (or some other standard method for solving an initial problem for ODEs, see, e.g.,~\cite{butcher2016numerical}). At the end of the integration interval $0\leq\Phi\leq \Delta$, one obtains
$t^{next}\equiv t(\Delta)$ and $\vec{X}(t^{next})\equiv \vec{X}(\Delta)$. Then, the simulation is continued as described in section~\ref{sec:fm} above.

Let us now consider a situation where the total rate $\Lambda$ can vanish at some time intervals. In this case, a global replacement of independent variables like in (\ref{eq:newx},\ref{eq:newt}) is not possible. In such a situation, we suggest a procedure that basically mimics the construction of the Poincar\'e map according to the H\'enon method~\cite{Henon-82}. One first integrates the system (\ref{eq:bode},\ref{eq:lambdaode}) using time $t$ as the independent variable and checks the condition $\Phi=\Delta$ at a sequence of small time intervals (say, at time instants $t_i$). When the interval $(t_{i-1},t_i)$ is found on which the variable $\Phi$ crosses the level $\Delta$ (i.e., $\Phi(t_{i-1})<\Delta<\Phi(t_i)$), then one makes an adjustment step. Namely, one integrates
the system (\ref{eq:newx},\ref{eq:newt}) from the initial condition $(\vec{X}(t_i),t_i)$ at $\Phi=\Phi(t_i)$ up to the desired value of the independent variable $\Phi=\Delta$. In most cases, it is sufficient to perform just one numerical integration step of length $\Delta-\Phi(t_i)$ (notice that this step is negative because at time $t_i$ one overshoots the level $\Phi=\Delta$). Equivalently, one can integrate from the initial condition $(\vec{X}(t_{i-1}),t_{i-1})$ at $\Phi=\Phi(t_{i-1})$ with a positive time step $\Delta-\Phi(t_{i-1})$. Still, this latter variant is slightly more complex in implementation because one must remember the states at the two last time steps. Because between the time instants $t_{i-1}$ and $t_i$ the variable
$\Phi$ grows, it means that on this interval $\Lambda>0$ and one can safely use equations (\ref{eq:newx},\ref{eq:newt}); here also the  smallness of intervals $t_{i}-t_{i-1}$ is important.

Several remarks are in order.

(1) The method suggested is not exact because it is based on a numerical solution of a system of ODEs. However, for such a problem, one can quite easily control accuracy. We mention that in typical cases, the evolution of the deterministic equations (\ref{eq:bode}) can be anyhow performed only numerically.

(2) The approach can also be applied to Markov processes without deterministic dynamics but with explicit time dependence of the rates. In this case, the equations for $\vec{X}$ are absent, but the problem (\ref{eq:lambdaode}) still has to be solved. We suggest using Eq.~(\ref{eq:newt}) instead.

(3) In our approach, we assumed sufficient smoothness of functions $\vec{F},\lambda_i$ on their arguments. This is also needed for an efficient application of numerical integration, where accuracy depends on the smoothness of the r.h.s. In particular, to have a well-defined system of ODEs  (\ref{eq:newx},\ref{eq:newt}), it is required that the time-dependence of rates can be explicitly calculated at any $t$. This does not allow for a stochastic dependence of the rates on time.

(4) The rates may be so small that the variable $t$ as a function of $\Phi$ in (\ref{eq:newt}) rapidly grows. In such cases, one should define an upper bound $t^{max}$ so that if in the course of integration $t>t^{max}$, the integration stops: there are no further jump events in the Markov process.

\section{Example: the Morris-Lecar system}
\label{sec:ml}

In this section, we illustrate the method by stochastic simulations of the Morris-Lecar model 
of neuron activity~\cite{morris1981voltage}. We have chosen this example because it has been studied with other methods in Refs.~\cite{anderson2015stochastic,lemaire2020thinning}. The Morris-Lecar model contains an equation for the membrane voltage and, generally, takes into account two types of conducting channels: calcium and potassium.
In the simplest formulation, calcium channels are treated in the mean-field approximation with corresponding
nonlinear terms in the voltage equation, while potassium channels are treated stochastically. We will explore this version below.

The equation for the time-dependent voltage $V(t)$ (this corresponds to the general continuous variable $\vec{X}$ above)
reads
\begin{equation}
\begin{gathered}
\frac{d V}{dt}=F(V,N_{open})=\\
\frac{1}{C}\left(I_{ext}-g_{Ca}m_\infty(V) (V-V_{Ca})-g_L(V-V_L)-g_K \frac{N_{open}}{N_K}(V-V_K)
\right)\;.
\end{gathered}
\label{eq:mlvolt}
\end{equation}
Here $N_{open}$ is a discrete variable counting the number of open potassium channels $0\leq N_{open}\leq N_K$ (this variable corresponds to the discrete variables $\vec{Y}$ above). The jump events are openings and closings of a channel. The rates for opening and closing of a single channel are $\alpha(V)$ and $\beta(V)$, respectively.
The total rate for opening at least one channel is $\lambda_1=\alpha(V)(N_K-N_{open})$, at this event $N_{open}
\to N_{open}+1$. The corresponding rate for closing is
$\lambda_2=\beta(V) N_{open}$, at this event $N_{open}\to N_{open}-1$.  The total rate for an opening/closing to occur is $\Lambda(V,N_{open})=\alpha(V)(N_K-N_{open})+\beta(V) N_{open}$.
The functions and constants in (\ref{eq:mlvolt}) are (we take the parameters from Ref.~\cite{anderson2015stochastic})
\begin{gather*}
m_\infty(V)=\frac{1}{2}\left(1+\tanh\left(\frac{V-V_a}{V_b}\right)\right)\;,\\
\alpha(V)=\phi\frac{\cosh (\xi/2)}{1+e^{-2\xi}}\;,\qquad
\beta(V)=\phi\frac{\cosh (\xi/2)}{1+e^{2\xi}}\;,\qquad \xi=\frac{V-V_c}{V_d}\;,\\
C=20\;,\qquad V_K=-84\;, \qquad V_L=-60\;, \qquad V_{Ca}=120\;,\\
I_{ext}=100\;,\qquad g_K=8\;,\qquad g_L=2\;,\qquad g_{Ca}=4.4\;,\\
V_a=-1.2\;,\qquad V_b=18\;,\qquad V_c=2\;,\qquad V_d=30\;,\qquad\phi=0.04\;.
\end{gather*}

At each step of the simulation, we used the following algorithm as described in Sect.~\ref{sec:ssm}:\\
1. Define a random number $\Delta=-\ln r_1$, where $r_1$ is uniform in $(0,1]$. \\
2. Integrate equations 
\begin{equation}
\begin{aligned}
\frac{d V}{d \Phi}&
=\frac{F(V,N_{open})}{\Lambda(V,N_{open})}\;,\\
\frac{dt}{d \Phi}&=\frac{1}{\Lambda(V,N_{open})}
\end{aligned}
\label{eq:mlinv}
\end{equation}
on the interval $0\leq \Phi\leq \Delta$ starting with $V^{last},t^{last}$ to find $V^{next},t^{next}$ at the end of the integration.\\
3. Generate a uniformly distributed random number $r_2$.\\
4. If $\displaystyle r_2<\frac{\alpha(V^{next}) (N-N_{open})}{\Lambda(V^{next},N_{open})}$ then $N_{open}\to N_{open}+1$, otherwise $N_{open}\to N_{open}-1$.\\
5. Replace $V^{next}\to V^{last}$ and $t^{next}\to t^{last}$ and repeat from step 1.
 
We integrated the system  (\ref{eq:mlinv}) using the Dormand-Prince-Runge-Kutta method~\cite{butcher2016numerical} with a fixed time step. First, we fixed, for each value of the total number of potassium channels $N_K$, a maximal integration step $h$. Then, the number of integration steps $L$ for each interval $0\leq \Phi\leq \Delta$ was determined as $L=\lfloor \Delta/h\rfloor+1$ where $\lfloor\cdot\rfloor$ is the integer part of a real number. The constant integration step is then $\Delta/L$.

\begin{figure}
\centering
\includegraphics[width=0.8\columnwidth]{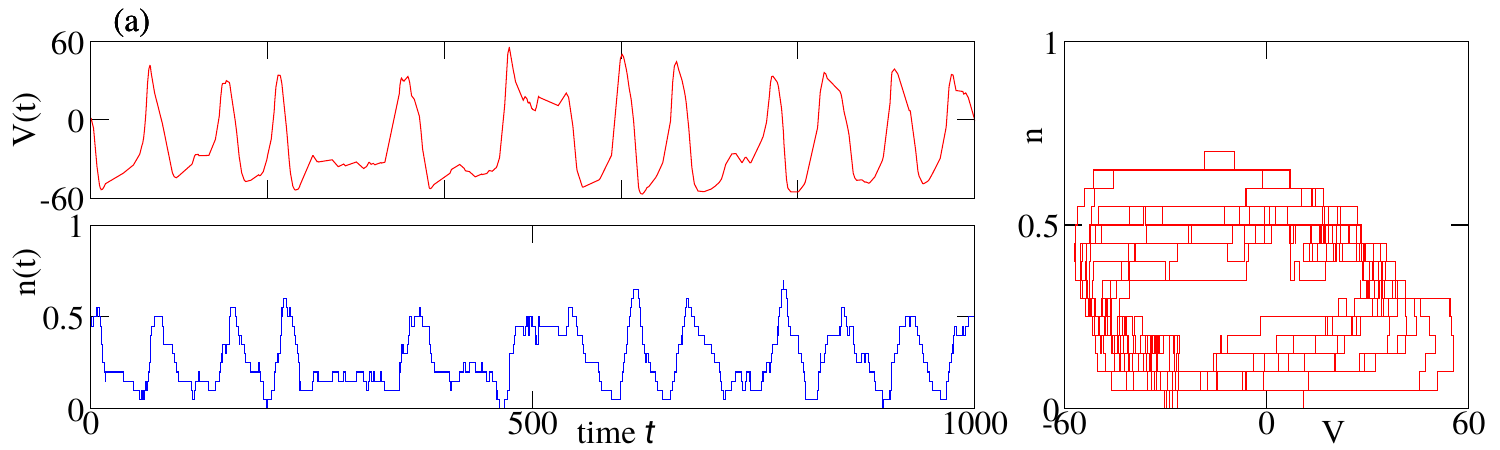}\\
\includegraphics[width=0.8\columnwidth]{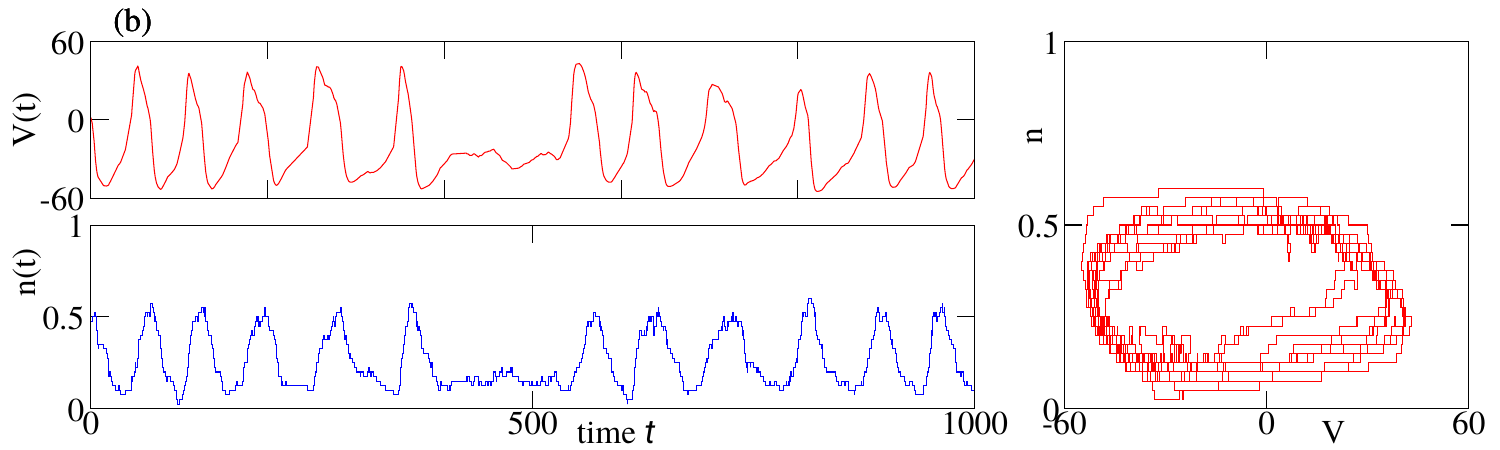}\\
\includegraphics[width=0.8\columnwidth]{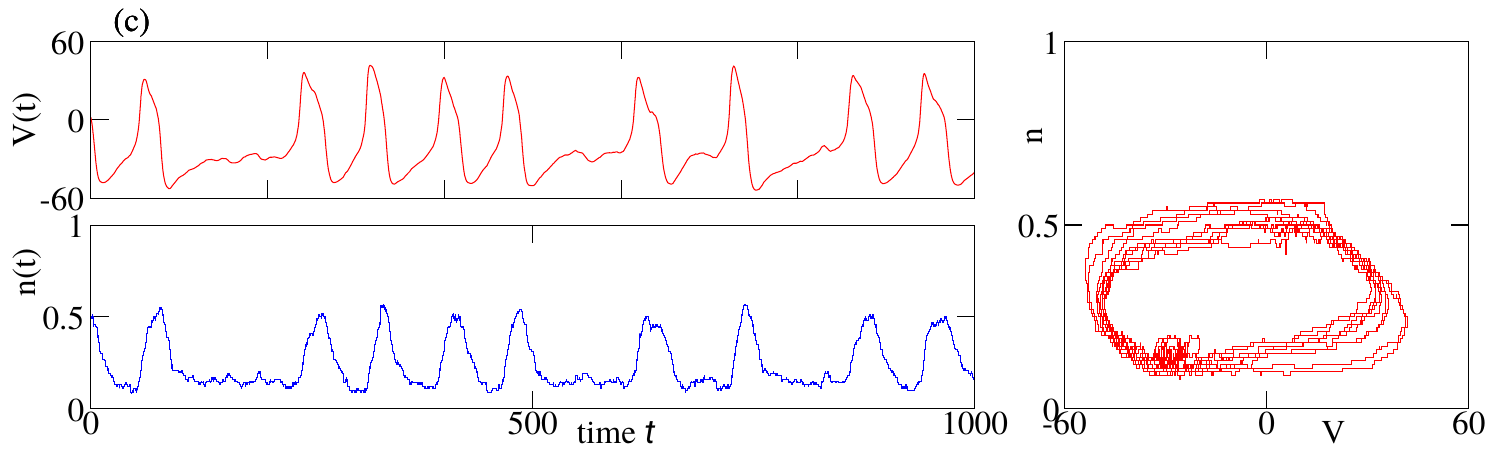}\\
\includegraphics[width=0.8\columnwidth]{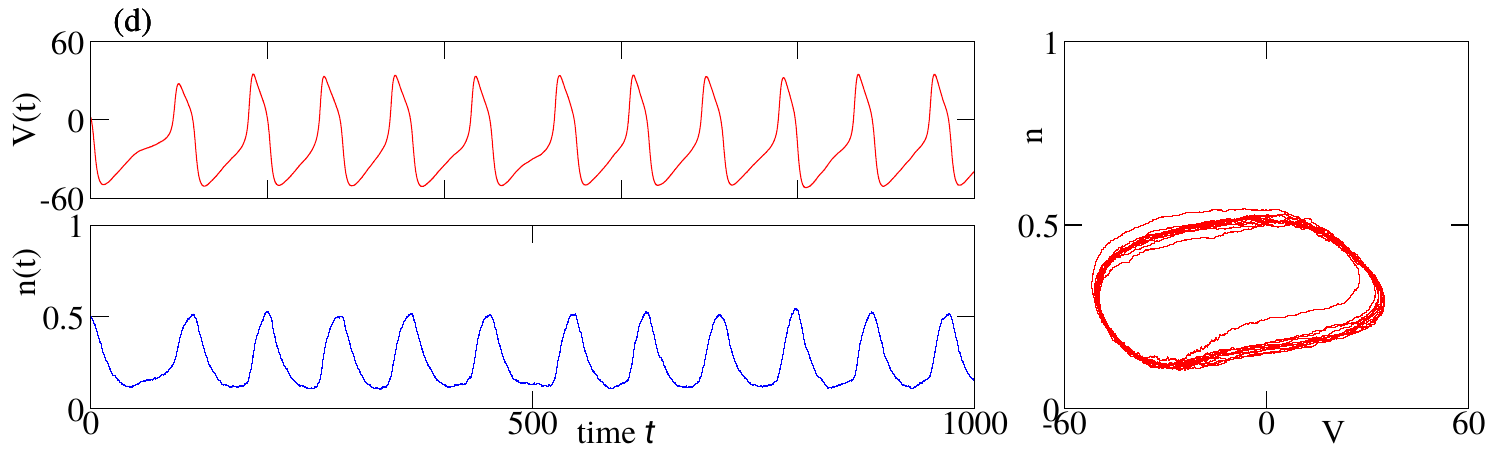}
\caption{Time series $V(t),n(t)$ of stochastic simulations of the Morris-Lecar model with different numbers of channels: (a) $N_K=20$; (b) $N_K=40$; (c) $N_K=100$; (d) $N_K=1000$. Right column shows the trajectories on the plane $(V,n)$.  }
\label{fig:traj}
\end{figure}

We present the obtained trajectories $ V (t), n (t) = N_{open} (t)/N_K$ in Fig.~\ref{fig:traj}. Here, we also show projections on the plane $(V,n)$. One can see that with an increase in the number of channels $N_K$, the stochasticity of channel openings and closings becomes less pronounced, and the time dependencies become effectively smooth. 

\begin{figure}[!htb]
\centering
\includegraphics[width=\columnwidth]{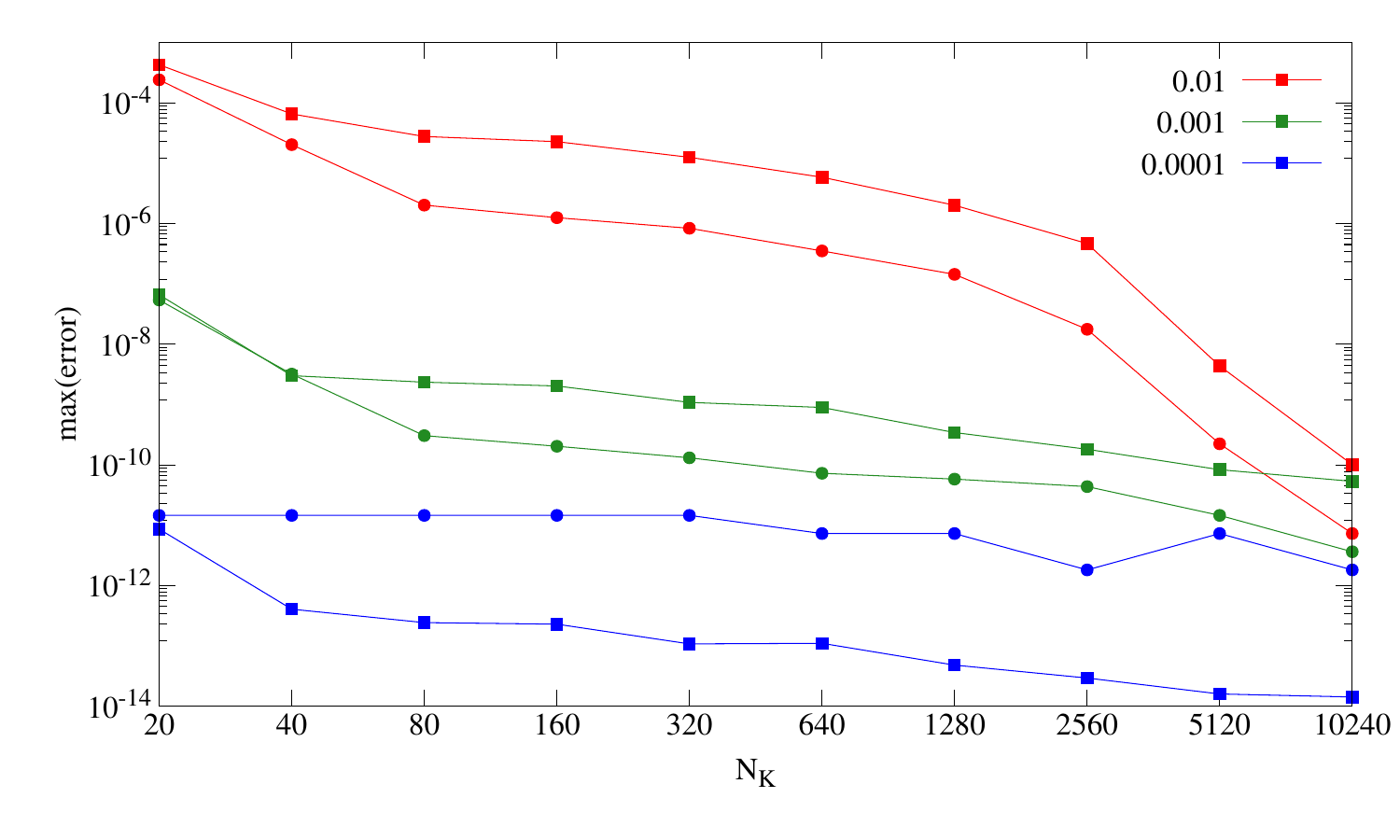}
\caption{Maximal errors for different values of $h_0$ in dependence on the number of channels $N_K$. Squares: errors for variable $V$; circles: errors for the times $t$. Red color: $h_0=0.01$; green color: $h_0=0.001$;
blue color: $h_0=0.0001$. }
\label{fig:acc}
\end{figure}

For the purpose of this paper, the most important thing is to analyze errors.  In the Dormand-Prince method, one can evaluate an error in every variable at every step. If several steps were needed for integration from $t^{last}$ to $t^{next}$, we summed up the absolute values of the errors at each small step to estimate the error of finding the next instant of time and the next voltage. Furthermore, we calculated the maximal values of these errors in the long run. The results are presented in Fig~\ref{fig:acc}. Because in Eqs. (\ref{eq:mlinv}) the r.h.s. are inversely proportional to the number of channels $N_K$, we adopted the following rule: The maximal integration step was chosen as $h=N_K h_0$. We present the results for three values of the constant $h_0$: $h_0=10^{-2},\;10^{-3},\; 10^{-4}$. Of course, smaller $h_0$ provides better accuracy, but even with $h_0=0.001$, good results can be obtained in the whole range of explored numbers of the channels $N_K$.

\begin{figure}[!htb]
\centering
\includegraphics[width=\columnwidth]{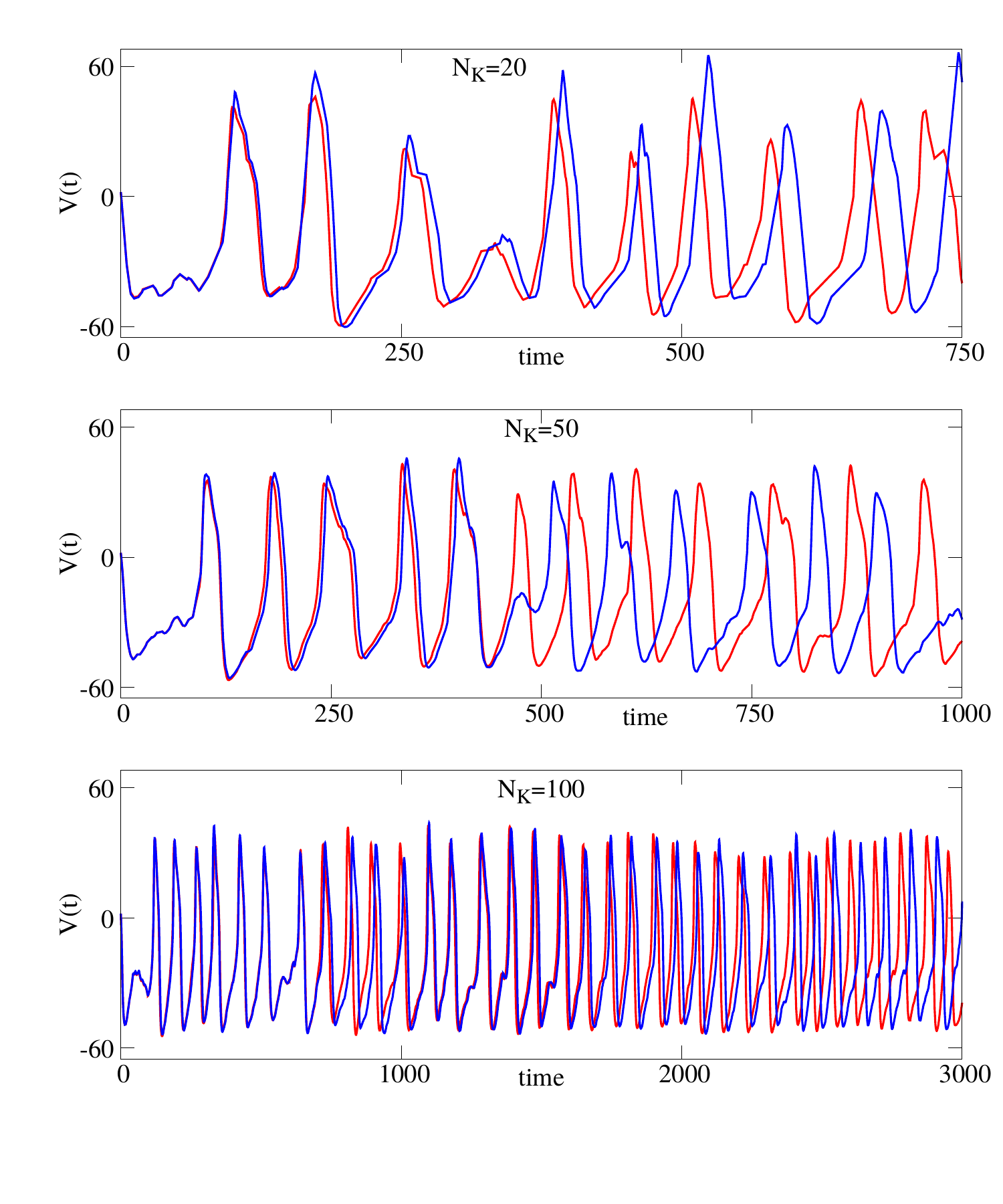}
\caption{Comparison of trajectories created with the same sequences of random numbers $r_1,r_2$ with the method described in Section~\ref{sec:ssm} (red curves) and with the approximate method where variations of the rates are neglected on the interval between the events (blue curves).
Numbers of channels are given on the panels. Notice the different time ranges of the panels. Integration was performed with the Dormand-Prince method with a fixed step $10^{-3}\cdot N_K$.}
\label{fig:comp}
\end{figure}

Finally, we compare the simulations using the method described in Section~\ref{sec:ssm} with approximate simulations (commonly adopted in the literature) where the time dependence of the rates within a step is neglected. 
Namely, in the integral in expression \eqref{eq:intcond}
one sets $\Lambda\approx\Lambda(\vec{X}(t^{last}),\vec{Y}(t^{last}),t^{last})$. Then the integral can be trivially calculated, and the time of the next event is $t^{next}\approx t^{last}+\Delta/\Lambda(\vec{X}(t^{last}),\vec{Y}(t^{last}),t^{last})$. One expects that such an approximation works well if the rate $\Lambda$ is large, so the time interval between the events is small (much smaller than the characteristic timescale of the rate variation). In the context of the Morris-Lecar system, this corresponds to a large number of channels $N_K$. To compare the two methods, we performed simulations of the Morris-Lecar system using the two methods described, with the same initial conditions and the same sequence of random numbers $r_1,r_2$. As a result, the initial stages of the trajectories presented in Fig.~\ref{fig:comp} coincide, but after some time interval (which, as expected, is longer for larger values of $N_K$), they diverge.

\section{Accuracy and Performance}
\label{sec:ap}
\begin{figure}[!htb]
\centering
\includegraphics[width=0.6\columnwidth]{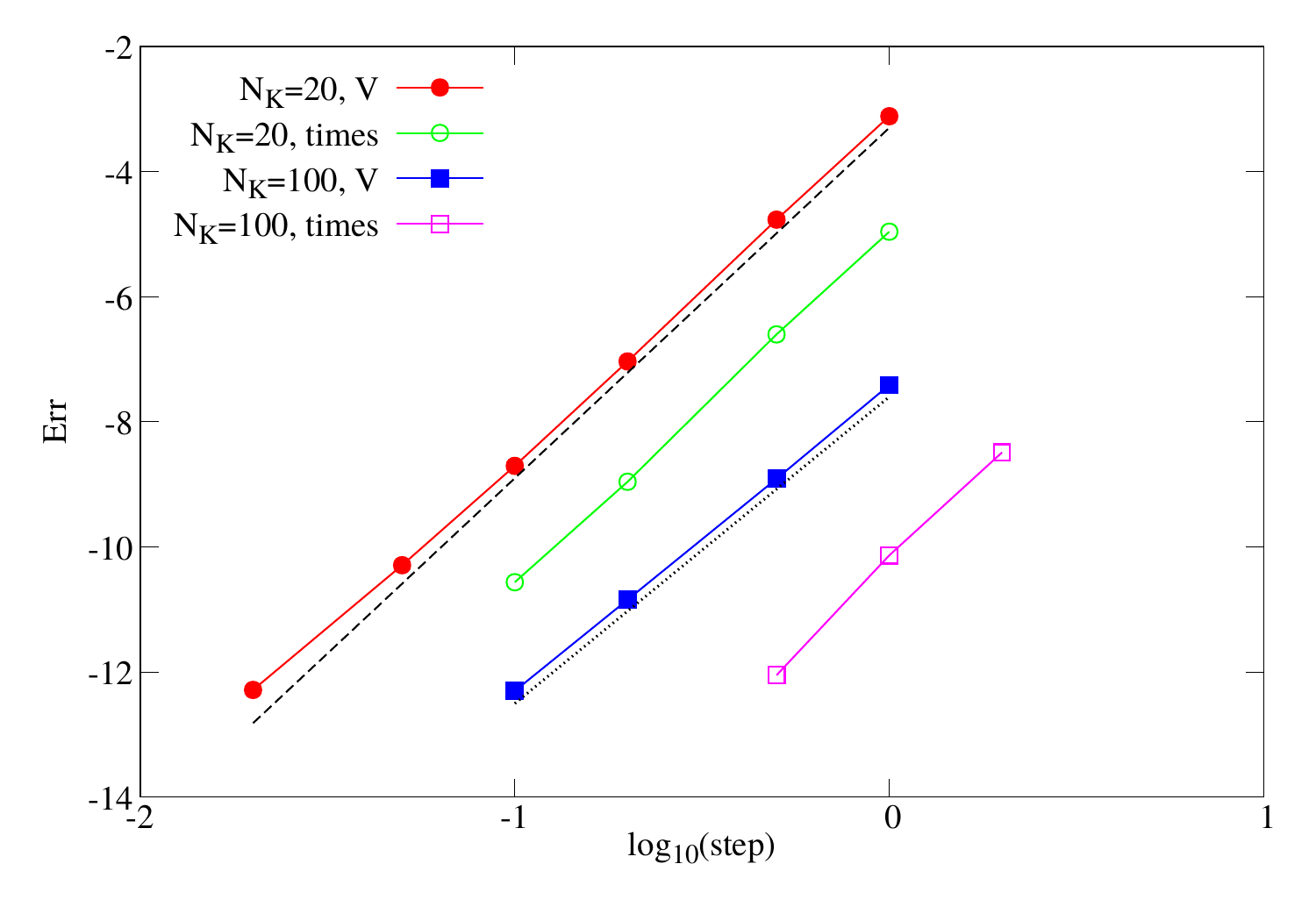}
\caption{Accuracies in voltages (filled markers) and time differences (open markers) for $N_K=20$ (circles) and $N_K=100$ (squares). The slope of the dashed line is $5.6$, the slope of the dotted line is $4.9$.}
\label{fig:acc2}
\end{figure}

In the suggested method, the only approximative numerical technique is approximation of the solution of the system of ODEs (\ref{eq:newx},\ref{eq:newt}) by means of one of the standard methods \cite{butcher2016numerical}. Thus, its convergence is assured by the convergence of the corresponding methods. To illustrate this, we calculated errors in simulations of the Morris-Lecar system for two values $N_k=20$ and $N_k=100$(cf.~Fig.~\ref{fig:comp}) for the Dormand-Prince method of numerical solution of the ODEs. First, we calculated a ``reference'' trajectory $\hat{V}_k,\{\hat{t}_k\}$ (here $\{\hat{t}_k\}$ are instants of time at which random openings/closing happen, and $\hat{V}_k$ are voltages at these times) using very small time steps ($10^{-4}$ for $N_K=20$ and $10^{-3}$ for $N_K=100$), at these steps the accuracy is limited by roundoff errors. The length of trajectory was set to $L=10^4$ random events of channel opening/closing. Next, we calculated, for the same initial conditions and for the same sequence of pseudo-random numbers, trajectories with larger integration steps. Because the dynamics is not chaotic (in fact, in the deterministic limit the dynamics is periodic), the trajectories do not differ much on the finite time interval. We evaluated the accuracy using two quantities. First, we calculated the error in the voltages as $Err(V)=L^{-1} \sum_{k=1}^L\log_{10}|V_k-\hat{V}_k|$. Next, we calculated the error in the inter-event time intervals as 
$Err(t)=(L-1)^{-1} \sum_{k=2}^L\log_{10}|(t_k-t_{k-1})-(\hat{t}_k-\hat{t}_{k-1})|$.
Additionally, we averaged over $100$ different realizations of sequences of generated pseudo-random numbers.
Both errors are depicted vs the integration step in Fig.~\ref{fig:acc2}. We note that larger ranges for these dependencies are hardly possible, because the errors are practically bounded from below at $\approx 10^{-12}$ due to roundoffs. In both cases the approximate scaling $Err\sim h^\kappa$ holds, with $\kappa\approx 5.6$ for $N_K=20$ and $\kappa\approx 4.9$ for $N_K=100$.

Next, we discuss other numerical methods for PDMP models. An alternative approach for exact simulation of Markov processes with time-dependent rates is the thinning method~\cite{Lewis-79,lemaire2018exact}. It requires explicit knowledge of the time-dependent total rate $\Lambda(t)$ and its upper bound $\overline{\Lambda}>\Lambda(t)$. Then, first  two random numbers 
$(\tilde{t},\tilde{\Lambda})$ are generated, and then it is checked whether $\tilde{\Lambda}<\Lambda(\tilde{t})$. If the condition is not fulfilled, the generated pair is rejected, and a new attempt is performed. This method can also be applied in cases where the ODE for the continuous variables $\vec{X}(t)$ can be solved explicitly, providing an explicit expression for the rates as functions of time. This is the case for the Hodgkin-Huxley model of neuron activity~\cite{hille2001}, where the equation for the time-dependent voltage is linear (contrary to nonlinear equation~(\ref{eq:mlvolt}) in the Morris-Lecar model). One can express its solution analytically and correspondingly apply the
 thinning method~\cite{lemaire2018exact}. The only remaining problem is to find a good estimate for the upper bound $\overline{\Lambda}$ to reduce the number of rejections. If the equations for the continuous variable $\vec{X}(t)$ are nonlinear and not exactly solvable, the exact time-dependence of the rates is unavailable. 
 In Ref.~\cite{lemaire2020thinning}, it was suggested to use Euler-type numerical integration to obtain an approximate time-dependence of the rates and to explore different integration steps to converge to a correct value of the time interval. In the same spirit is the approach of Ref.~\cite{ding2016numerical}, where a piecewise-exponential approximation of the solution of the ODE is suggested. We believe that our method is more straightforward in implementation.

Below we compare the performance of our approach based on the solution of Eqs.~ (\ref{eq:newx},\ref{eq:newt}) with the methods based on the solution of Eqs.~\eqref{eq:bode},\eqref{eq:lambdaode} and finding the time instant where $\Phi=\Delta$. Such simulations have been performed in Refs.~\cite{riedler2013almost,anderson2015stochastic,ding2016numerical}. The authors of ~\cite{riedler2013almost,anderson2015stochastic} used MATLAB\textsuperscript{\textregistered}'s \texttt{ode45} routine with a built-in event detection feature. In \cite{ding2016numerical}
it is suggested to solve  Eqs.~\eqref{eq:bode},\eqref{eq:lambdaode} with some time step $h$ up to the time instant $t_{next}$ at which the threshold is overshooted 
$\Phi(t_{next})>\Delta$. Then, the event time should be found from a linear interpolation of $\Phi(t)$ on the last interval $(t_{next}-h,t_{next})$. A more accurate event detection can be achieved if the linear interpolation is successively applied  $m>1$ times. The accuracy for such a method depends both on the accuracy of the numerical integration of the ODEs and on the accuracy of the numerical solution of the condition $\Phi(t)=\Delta$, thus it makes no sense to use large $m$ because then the total accuracy will be limited by the ODE solver.

\begin{figure}[!htb]
\centering
\includegraphics[width=0.48\columnwidth]{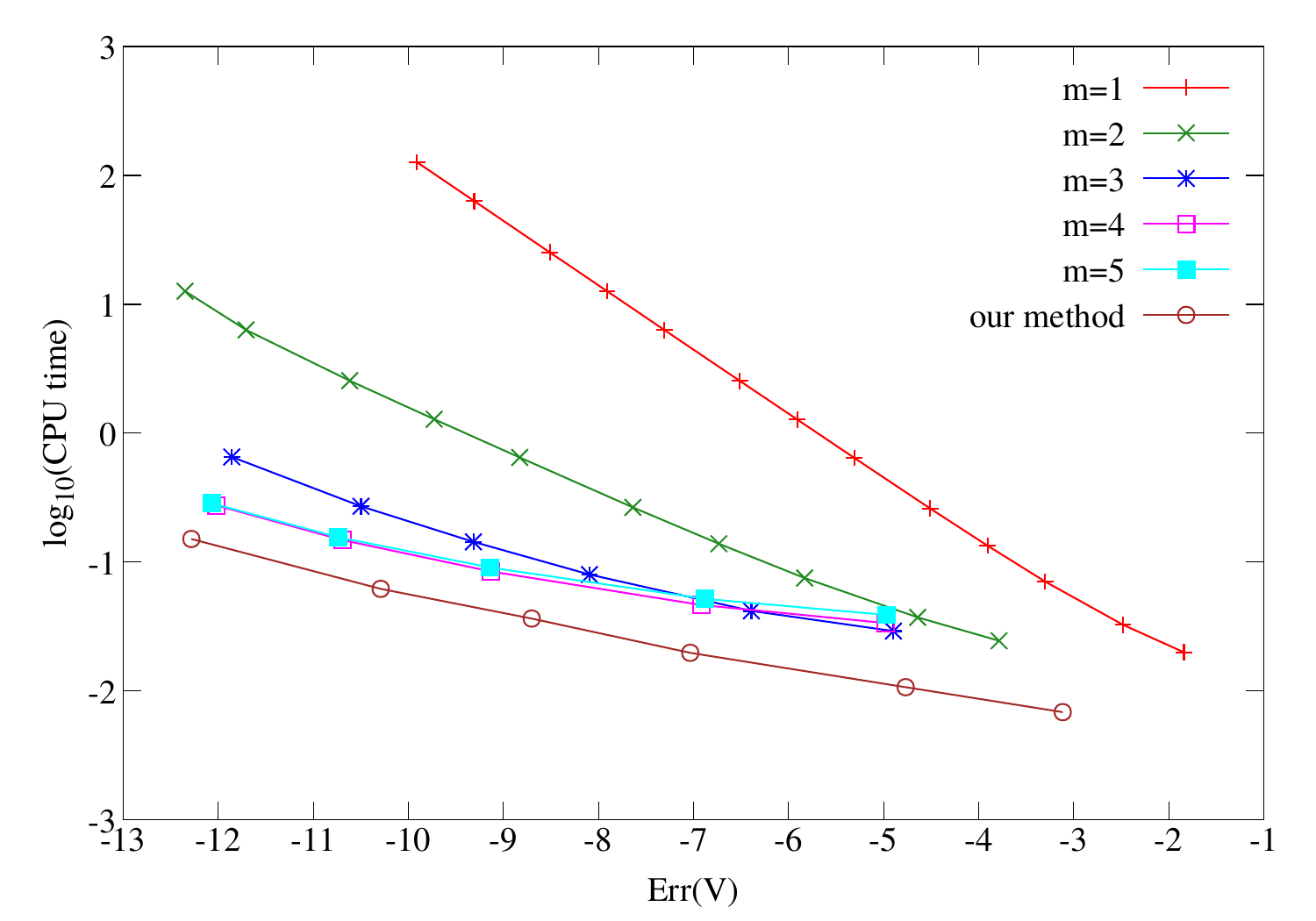}\hfill
\includegraphics[width=0.48\columnwidth]{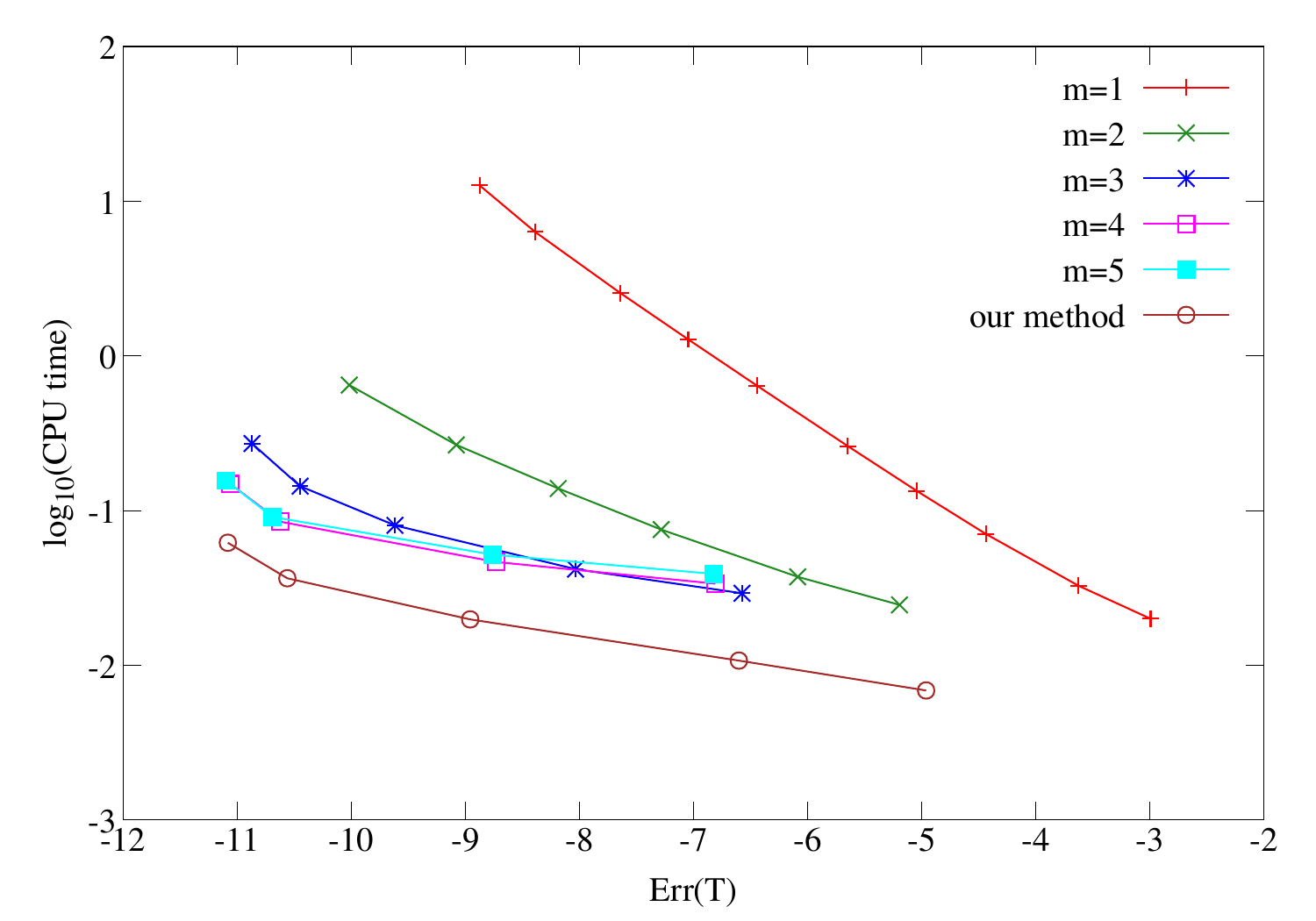}
\caption{Average CPU times vs accuracies $Err(V)$ (left panel) and $Err(t)$ (right panel) for our method and the methods based on the solution of ODEs Eqs.~\eqref{eq:bode},\eqref{eq:lambdaode}, for different numbers of iteration steps $m$. The errors are defined as the averaged decimal logarithms, eo, e.g., the value $Err=-8$ corresponds to the absolute error $\approx 10^{-8}$. }
\label{fig:cpu}
\end{figure}

To compare different methods, we used in all cases the Dormand-Prince solver of the ODEs, and implemented the event detection using successive linear approximations with $1\leq m\leq 5$ as described above. The results for the case $N_K=20$ are presented in Fig.~\ref{fig:cpu}. Here we used the same definition of the averaged errors as described above in this section, additionally we averaged the decimal logarithms of the elapsed CPU times.  First, one can see that the accuracy and performance of event-location methods does increase significantly with $m$ for $m>4$. In all cases our method overperforms the event-location methods by a factor $\approx 2$. Additionally, it is easier in implementation.

\section{Conclusion}
\label{sec:concl}

In summary, we have suggested an approach that significantly accelerates stochastic modeling of piecewise-deterministic Markov process. The main idea is to solve the continuous-time ordinary differential equations between the jump events using the cumulative total rate as an independent variable. In this formulation, the integration interval is fixed (to a random number having exponential distribution). Thus, a numerical implementation of such integration is straightforward and does not require additional steps. Furthermore, accuracy can be easily controlled because standard numerical methods like the Dormand-Prince version of the Runge-Kutta method can be used. We have illustrated our approach by simulating the stochastic version of the Morris-Lecar model of neuron activity, where potassium channels are randomly closed and opened.

Finally, we mention that together with the Gillespie Direct Method explored in this paper, there are its variants 
called first-event and next-event methods (see~\cite{masuda2022gillespie} for a recent review). There, the same numerical bottleneck is finding the time instant of the next event from a time- and variable-dependent rate, and our approach also works for these variants.

\backmatter
\bmhead{Acknowledgements}
I thank A. Deser for fruitful discussions.


%
%



\end{document}